\documentclass[journal=jpccck,
               manuscript=article,
               layout=twocolumn,
               ]{achemso}
\setkeys{acs}{articletitle=true}
\usepackage{cancel}
\usepackage{amsmath}
\usepackage{physics}
\usepackage{graphicx}
\usepackage{amssymb}
\usepackage{amsthm}
\usepackage{bm}
\usepackage{threeparttable} 
\usepackage{dcolumn}
\newcolumntype{d}[1]{D{.}{.}{#1}}
\usepackage{braket}
\usepackage{booktabs}

\usepackage{ragged2e}
\usepackage{txfonts}

\makeatletter
\let\l@addto@macro\relax
\makeatother
\usepackage[fontsize=9pt]{scrextend}

\usepackage{color}
\usepackage[breaklinks=true,colorlinks=true,allcolors=blue]{hyperref}
\usepackage{cleveref}

\let\vr\undefined
\newcommand{\vr}{{\bm{r}}}

\title{Excitons and their Fine Structure in Lead Halide Perovskite Nanocrystals from Atomistic GW/BSE Calculations}

\author{Giulia Biffi}
\affiliation{Istituto Italiano di Tecnologia, Via Morego 30, 16163 Genova, Italy}
\alsoaffiliation{Dipartimento di Chimica e Chimica Industriale, Università degli Studi di Genova, Via Dodecaneso, 31, 16146 Genova, Italy}
\altaffiliation{These authors contributed equally}
\author{Yeongsu Cho}
\affiliation{Department of Chemistry, Columbia University, New York, New York 10027 USA}
\altaffiliation{These authors contributed equally}
\author{Roman Krahne}
\affiliation{Istituto Italiano di Tecnologia, Via Morego 30, 16163 Genova, Italy}
\author{Timothy C. Berkelbach}
\email{t.berkelbach@columbia.edu}
\affiliation{Department of Chemistry, Columbia University, New York, New York 10027 USA}
\alsoaffiliation{Center for Computational Quantum Physics, Flatiron Institute, New York, New York 10010 USA}

\begin{document}


\maketitle

\begin{abstract}
Atomistically detailed computational studies of nanocrystals, such as those
derived from the promising lead-halide perovskites, are challenging due to the
large number of atoms and lack of symmetries to exploit.  Here, focusing on
methylammonium lead iodide nanocrystals, we combine a real-space tight binding
model with the GW approximation to the self-energy and obtain exciton
wavefunctions and absorption spectra via solutions of the associated
Bethe-Salpeter equation.  We find that the size dependence of carrier
confinement, dielectric contrast, electron-hole exchange, and exciton binding
energies has a strong impact on the lowest excitation energy, which can be tuned
by almost 1~eV over the diameter range of 2--6~nm.  Our calculated excitation
energies are about 0.2~eV higher than experimentally measured photoluminescence,
and they display the same qualitative size dependence.  Focusing on the fine
structure of the band-edge excitons, we find that the lowest-lying exciton is
spectroscopically dark and about 20--30~meV lower in energy than the
higher-lying triplet of bright states, whose degeneracy is slightly broken by
crystal field effects.
\end{abstract}

\section*{Introduction}

Lead-halide perovskite nanocrystals (LHP NCs), with formula APbX$_3$ where A is
a monovalent cation and X is a halide anion, exhibit remarkable electronic and optical
properties~\cite{Kovalenko2017,Shamsi2019,Dey2021}, suggesting promising
applications in
optoelectronics~\cite{Tan2014,Huang2016,Lee2018,Palei2018,Akkerman2018},
photonics~\cite{Sutherland2016,Caligiuri2018}, and
spintronics~\cite{Shrivastava2020}.
Compared to bulk LHPs, LHP NCs display anomalously short radiative lifetimes and
concomitant high photoluminescence quantum yields, the origin of which is still
under debate~\cite{Kovalenko2017,Imran2018,Shamsi2019,Dey2021}.  These interesting optical
properties and their temperature dependence have focused attention on the
exciton fine structure, i.e., the energy ordering and character of the band-edge
excitons~\cite{Fu2017,Becker2018,Sercel2019JCP,Sercel2019NL,Tamarat2019,Rossi2020,Rossi2020a,Dyksik2021}.
For example, the lowest-energy exciton of Cs-LHP NCs has been
suggested to be a bright (emissive) state with total angular momentum $J=1$,
which could explain the high photoluminescence quantum yield~\cite{Becker2018,
Sercel2019JCP, Sercel2019NL}.  This property would be in stark contrast with the
case of conventional organic and inorganic semiconductors, where the
lowest-energy exciton is a dark (non-emissive) state.  Other works have
suggested that this conventional behavior persists in the LHP NCs (i.e., the
lowest-energy exciton is a dark state with total angular momentum $J=0$) but
that exciton relaxation by phonons is suppressed at low temperatures, yielding
long lifetimes for the bright exciton in both NCs~\cite{Tamarat2019,Rossi2020,Rossi2020a}
and 2D LHPs~\cite{Dyksik2021}.

The properties of the band-edge excitons---whose energy separations are very
small, of the order of 1--10~meV---are influenced by quantum confinement,
electron-hole attraction and exchange~\cite{Becker2018, Sercel2019JCP,
Sercel2019NL, BenAich2019}, dielectric environment~\cite{Katan2019}, and lattice
structure (including the bulk and surface Rashba
effects)~\cite{Amat2014,Motta2015, Mehdizadeh2019, Saleh2021}, making precise
theoretical predictions an incredible challenge. This complexity has led to the
development of exciton models based on effective-mass or $\bm{k}\cdot\bm{p}$
Hamiltonians, pioneered especially by Efros and co-workers~\cite{Becker2018,
Sercel2019NL, Sercel2019JCP}, among others~\cite{BenAich2019, Blundell2022}.
Atomistic simulation, although desirable, is frustrated by the sizes of even the
smallest experimentally accessible NCs, which can have hundreds or thousands of
atoms.  Although atomistic force-fields or density functional theory can be
applied to study the ground-state and structural properties of systems of this
size~\cite{Buin, TenBrinck2016,TenBrinck2019,Perez2021}, accurate excited-state
theories are more expensive and generally inaccessible in a fully \textit{ab
initio} setting.

Here, we overcome this challenge and construct an atomistic orbital-dependent
tight-binding model parametrized by first-principles density functional theory
(DFT) calculations, after which we use a model dielectric function to apply
self-energy corrections via the GW
approximation~\cite{Hedin1965,Strinati1980,Hybertsen1986} and calculate the
energies and properties of excitons via the Bethe-Salpeter
equation~\cite{Albrecht1998,Benedict1998,Rohlfing1998}.  The computational
approach is similar to previous work by two of us on the properties of layered
quasi-two-dimensional LHPs~\cite{Cho2019}.  As a specific example, we study
methylammonium lead iodide (MAPbI$_3$) NCs, calculating the exciton binding
energies, absorption spectra, and band-edge fine structure.  Within the
approximations of our approach, we find that the lowest-energy exciton is always
a dark state.

We study MAPbI$_3$ for several reasons, although our approach is general and
could be applied to any LHP NC.  First, this LHP is one of the most studied in
its bulk form, especially for photovoltaics, partly due to its strong absorption
at low energies that facilitates sensitization~\cite{Kojima2009, Im2011}.
Moreover, bulk MAPbI$_3$ crystals have low lasing thresholds\cite{Cadelano2015}
and are generally more stable than Cs-based ones\cite{Xiao2017}.  For these
reasons, the atomic and electronic structure of bulk MAPbI$_3$ has been
extensively studied using DFT~\cite{Filippetti2015,Frohna2018} and the GW
approximation~\cite{Umari2014,Brivio2014,Filip2014,Filip2015}, providing
important points of comparison.  These valuable properties have motivated
experimental studies of MAPbX$_3$ NCs with high photoluminescence quantum yields
and controllable size, leading to tunable band gaps.  Such MAPbX$_3$ NCs have
been realized by colloidal
synthesis~\cite{Schmidt2014,Zhang2015,Huang2015,Xiao2019,Zhang2020, Ijaz2020}
and templated growth inside porous oxide films~\cite{Anaya2017,Rubino2021},
providing experimental results to which we can directly compare.

\section*{Computational Methods}

\begin{figure}[t] 
\centering
\includegraphics[scale=0.9]{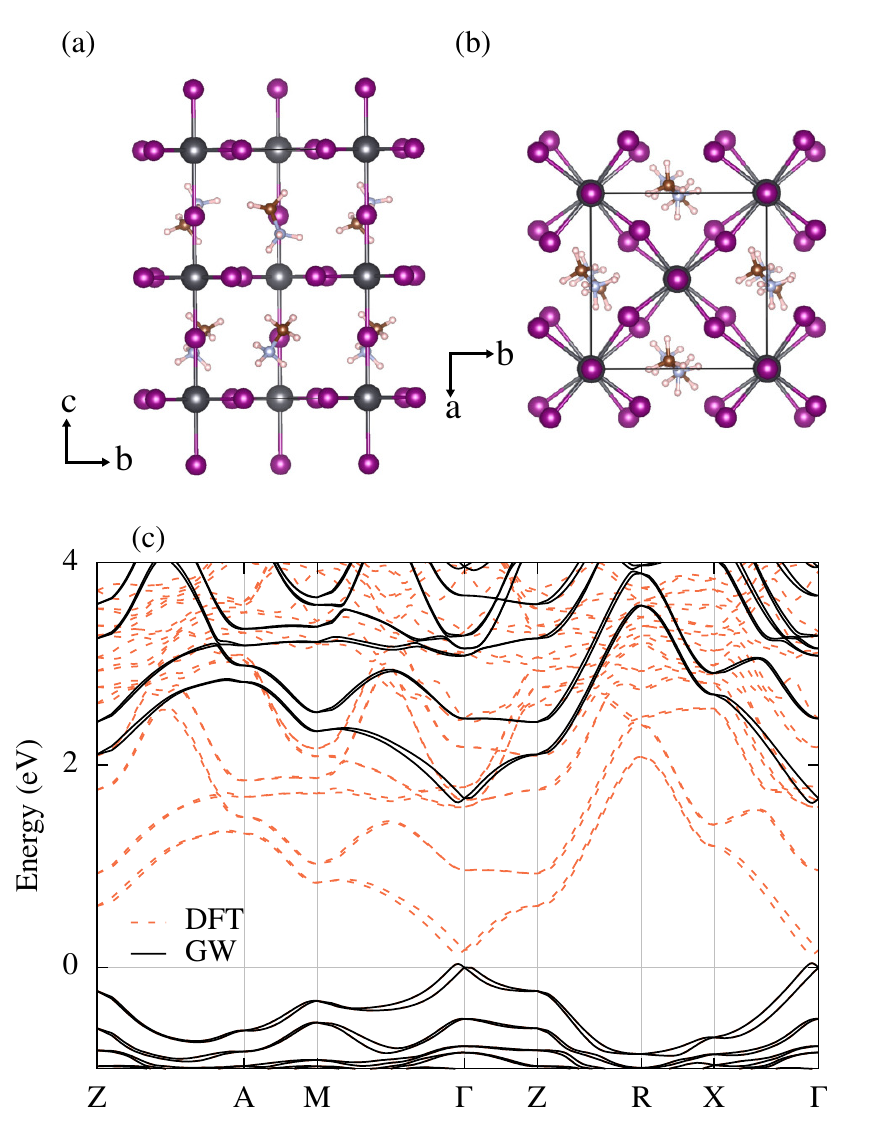} 
\caption{(a,b) Crystal structure of bulk tetragonal MAPbI$_3$ from
Ref.~\citenum{Frohna2018} and (c) band structure calculated with DFT (PBE with
spin-orbit coupling) and rigidly shifted to match the GW band
gap~\cite{Umari2014}. 
}
\label{fig:band}
\end{figure}

The parameters of our orbital-dependent tight-binding model are determined from
a DFT calculation of bulk MAPbI$_3$, which first requires the determination of
the crystal structure.  (To our knowledge, the crystal structure of LHP NCs is
still poorly understood, but the crystal structure of MAPbI$_3$ NCs is likely
tetragonal~\cite{Kovalenko2017}.  However, this assumption may be incorrect due
to surface effects and/or metastability.) From experiments, it is known that the
bulk MAPbI$_3$ exists in an orthorhombic phase below about 160~K, a tetragonal
phase between 160~K and 330~K, and a cubic phase above
330~K~\cite{Whitfield2016}. The computational prediction of crystal structures
in finite temperature phases is nontrivial because of anharmonic effects.  In
Ref.~\citenum{Frohna2018}, it was shown that using zero-temperature DFT for the
geometry optimization of the tetragonal and cubic phases yields a large
inversion symmetry breaking via distortion of the PbI framework, leading to a
large bulk Rashba effect. However, these crystal structures disagree with second
harmonic generation rotational anisotropy experiments.  Therefore, with interest
in room-temperature behavior, we reuse the structure proposed in
Ref.~\citenum{Frohna2018}, which was generated by fixing the PbI framework of
the tetragonal phase and relaxing only the MA cations, whose orientations yield
a small but nonzero Rashba effect.  This crystal structure is shown in
Fig.~\ref{fig:band}(a). 

With our MA-relaxed crystal structure of bulk MAPbI$_3$, we perform a DFT
calculation including spin-orbit coupling, using Quantum
Espresso~\cite{Giannozzi2009} with the PBE exchange-correlation
functional~\cite{Perdew1996}, relativistic PAW pseudopotentials, a kinetic
energy cutoff of 60~Ry, and a $6\times 6\times 6$ $k$-point mesh.  Using
wannier90~\cite{Mostofi2008}, the DFT solution is used to construct maximally
localized Wannier functions (MLWFs) $\phi_\mu(\vr)$, corresponding to I 5s and
5p orbitals and Pb 6s and 6p orbitals, as well as their tight-binding
Hamiltonian matrix elements $h_{\mu\nu}$ and dipole matrix elements
$r_{\mu\nu}^{e}$, $e\in\{x,y,z\}$.  Greek indices $\mu,\nu,\kappa,\lambda$ will
be used throughout to indicate MLWF atomic orbitals, including spin.  These
real-space tight-binding parameters are then used to build a mean-field
Hamiltonian $\mathbf{h}^\mathrm{NC}$ of the aperiodic NCs.  Our NC surfaces are
terminated with halide atoms, in agreement with experimental
observations~\cite{Zhang2015}, with exposed facets that are qualitatively
equivalent to the (100) surface of a cubic perovskite.  We use the same
tight-binding parameters for all atoms, even those near the surface of the NCs,
which can be understood as an approximate passivation that does not require
microscopic specification.  However, the details of surface passivation are
important for controlling trap states and associated non-radiative
recombination~\cite{TenBrinck2016,Xiao2019,Ijaz2020}.  Similarly, relaxation of
the atoms at the surface, which we neglect here, would modify these matrix
elements and the resulting electronic structure~\cite{TenBrinck2016,Saleh2021}.

With this tight-binding Hamiltonian matrix in the basis of MWLFs
$\phi_\mu(\vr)$, we calculate the molecular orbitals $C_{\mu p}$ and orbital
energies $\varepsilon_p$, $\mathbf{h}^\mathrm{NC}\mathbf{C} =
\mathbf{C}\bm{\varepsilon}$.  To the orbital energies, we add two self-energy
corrections in the spirit of the many-body GW approximation (at the level of
approximation made throughout this work, we do not distinguish degrees of
self-consistency in the self-energy, but our calculations are all performed in
the one-shot G$_0$W$_0$ approximation as described here).  The first correction
is a rigid shift of the conduction and valence band energies by $\pm \Delta/2$
to correct the bulk band gap to the previously calculated value of
1.67~eV~\cite{Umari2014}.  Our DFT and GW band structures of bulk tetragonal
MAPbI$_3$ are shown in Fig.~\ref{fig:band}, which can be seen to contain a small
but nonzero Rashba splitting at the band edges.
As is well-known~\cite{Umari2014,Brivio2014,Filip2014,Filip2015}, the GW
correction to the band gap of LHPs is sizable---about 1.5~eV in our case.  Our
second self-energy correction within the GW approximation uses a model
dielectric function and a static screening approximation to account for
dielectric constrast at the NC interface~\cite{Brus1983,Brus1984}, in the style
of a G$\delta$W calculation~\cite{Neaton2006}.  Combining these two corrections,
we calculate the GW quasiparticle energies of all conduction ($c$) and valence
($v$) bands,
\begin{subequations}
\begin{align}
E_c &= \varepsilon_c + \Delta/2 + \sum_\mu |C_{\mu c}|^2 \delta\Sigma(\vr_{\mu}) \\
E_v &= \varepsilon_v - \Delta/2 - \sum_\mu |C_{\mu v}|^2 \delta\Sigma(\vr_{\mu})
\end{align}
\end{subequations}
where $\vr_\mu$ is the position of the atom to which the MLWF $\phi_\mu(\vr)$ belongs,
\begin{equation}
\delta\Sigma(\vr_1) = \frac{1}{2}\lim_{\vr_2 \rightarrow \vr_1} 
    \left[W(\vr_1,\vr_2)-\frac{1}{\epsilon_\mathrm{p} |\vr_1-\vr_2|}\right],
\end{equation}
and $W(\vr_1,\vr_2)$ is the classical Coulomb interaction energy between two
charges in the NC at $\vr_1,\vr_2$~\cite{Brus1983}.  We use the analytical
Coulomb energy of charges in a dielectric cuboid with side lengths
$L_x,L_y,L_z$~\cite{Yang2002}, 
\begin{subequations}
\begin{align}
    W(\vr_1,\vr_2) &= \sum_{i,j,k=-\infty}^{\infty} \frac{[(\epsilon_\mathrm{p}-\epsilon_\mathrm{env})
        /(\epsilon_\mathrm{p}+\epsilon_\mathrm{env})]^{|i|+|j|+|k|}}
        {\epsilon_\mathrm{p}|\vr_1-\vr_2^{ijk}|} \\
    \vr_2^{ijk} &\equiv [(-1)^ix_2+iL_x, (-1)^jy_2+jL_y, (-1)^k z_2+kL_z].
\end{align}
\end{subequations}
With the surface termination described above, For a NC made of $n^3$ octahedra,
we use
$L_x=L_y = n\times(6.4~\AA)+2\times(1.5~\AA)$ and 
$L_z=n\times(6.5~\AA)+2\times(1.5~\AA)$, where 6.4~\AA\ and 6.5~\AA\ correspond
to the diagonals of the tetragonal octahedra and we have added an additional
dielectric buffer region of 1.5~\AA\ on all sides.  Throughout this work, we use
the bulk MAPbI$_3$ perovskite high-frequency dielectric constant
$\epsilon_\mathrm{p} = 6.1$ and an environmental dielectric constant of
$\varepsilon_\mathrm{env}=2.1$, representative of a generic low-dielectric
environment, such as organic ligands~\cite{Dirin2016,Akkerman2018,Shamsi2019}.
(For large NCs with small exciton binding energies that are on the order of
optical phonon frequencies, phonon screening suggests a larger value of
$\epsilon_\mathrm{p}$ or explicit treatment of electron-phonon
interactions~\cite{Filip2021}.)

To evaluate the optical and excitonic properties, we calculate the exciton
wavefunction as a solution to the Bethe-Salpeter equation (BSE) in the
Tamm-Dancoff approximation,
\begin{subequations}
\begin{align}
|\Psi^N\rangle &= \sum_{cv} A_{cv}^N \hat{a}^\dagger_c \hat{a}_v |0\rangle \\
E_N A_{cv}^N &= (E_c-E_v)A_{cv}^N + \sum_{c'v'}\left[(cv|V|v'c')-(cc'|W|v'v)\right] A_{c'v'}^N,
\end{align}
\end{subequations}
where $V$ and $W$ indicate the bare and screened Coulomb interactions and we use
$(11|22)$ notation for two-electron integrals.  For all results in the next
section, we included 200 valence bands and 200 conduction bands in the BSE
calculation.  In approximate theories of Wannier-type excitons, the exchange
interaction is commonly neglected on the basis of its short range, but as it is
critical in the determination of the exciton fine structure~\cite{BenAich2019},
we include it here.  The bare exchange integral is approximated as
\begin{equation}
(cv|V|v'c') = \sum_{\mu\kappa}
    C_{\mu c}^* C_{\mu v} C_{\kappa v'}^* C_{\kappa c'} V_{\mu\kappa} 
    + \sum_{\mu\neq\kappa}^\mathrm{intra} 
    C_{\mu c}^* C_{\kappa v} C_{\kappa v'}^* C_{\mu c'}
    (\mu\kappa|\kappa\mu)
\end{equation}
where $V_{\mu\kappa} = |\vr_\mu-\vr_\kappa|^{-1}$ if $\mu$ and $\kappa$ are not
on the same atom and $V_{\mu\kappa} = (\mu\mu|\kappa\kappa)$ otherwise.  Note
that our treatment includes both short-range and long-range exchange effects.
The screened Coulomb interaction is approximated similarly but without
intra-atomic exchange integrals, 
\begin{equation}
(cc'|W|v'v) = \sum_{\mu\kappa}
    C_{\mu c}^* C_{\mu c'} C_{\kappa v'}^* C_{\kappa v} W_{\mu\kappa} 
\end{equation}
where $W_{\mu\kappa} = W(\vr_\mu,\vr_\kappa)$ if $\mu$ and $\kappa$ are not on the same atom
and $W_{\mu\kappa} = (\mu\mu|\kappa\kappa)$ otherwise (i.e., the Coulomb interaction
is unscreened at the intra-atomic length scale).  The intra-atomic repulsion
integrals $(\mu\mu|\kappa\kappa)$ and $(\mu\kappa|\kappa\mu)$ for Pb and I s and
p orbitals are calculated using PySCF~\cite{Sun2017,Sun2020} with the def2-TZVP
basis set~\cite{Weigend2005}. 

From the BSE exciton eigenvectors $A_{cv}^{N}$, we calculate the polarized
absorption spectrum and its average
\begin{subequations}
\begin{align}
I_e(\omega) &\propto \sum_N
    \Big| \sum_{cv} r_{cv}^{e} A_{cv}^N \Big|^2 \delta(\hbar\omega-E_N)  \\
I(\omega) &= \sum_{e\in\{x,y,z\}} I_e(\omega).
\end{align}
\end{subequations}
Because of our nonperturbative inclusion of spin-orbit coupling, the molecular
orbitals have mixed spin character, i.e., $m_s$ is not a good quantum number.
Therefore, the total spin $S$ is not a good quantum number for the exciton
wavefunctions and they cannot be separated into spin singlets ($S=0$) and
triplets ($S=1$) with familiar dipole selection rules.  Rather, the spectral
intensity of the excited states is determined by the lattice symmetry in
combination with the spin and orbital character of the molecular orbitals
participating in the dominant transitions.  This feature is one of the primary
origins of the complex fine structure of low-lying excitons in lead-halide
perovskites.

\section*{Results and Discussion}

\begin{figure}[t] 
\centering
\includegraphics[scale=0.9]{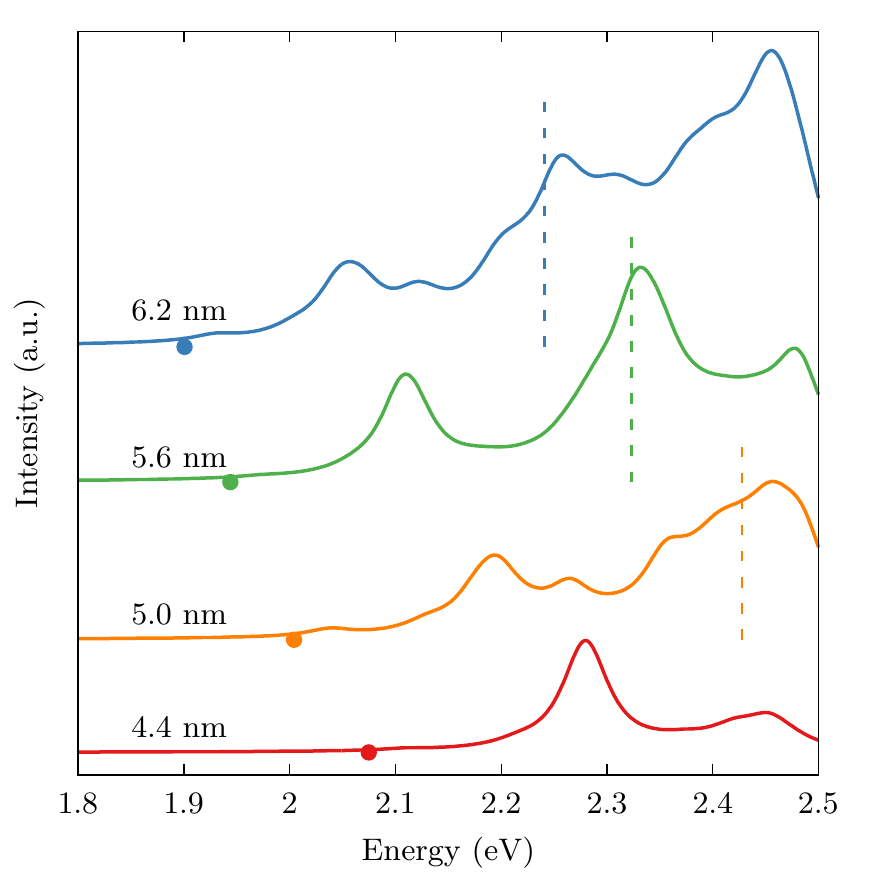} 
\caption{Absorption spectra of  MAPbI$_3$ nanocrystals with side length
$L=4.4$--$6.2$~nm ($n=7-10$). The first excitation, always spectroscopically dark, 
is indicated by a filled circle, while the band gap is indicated by vertical dashed line.}
\label{fig:spec}
\end{figure}

We have performed calculations on halide-terminated, cube-shaped NCs containing
$n^3$ octahedra, with $n=3$--$10$, corresponding to a NC edge length 
$L \approx 1.9$--$6.2$~nm; our largest NCs include almost 4300 Pb and I atoms.
Assuming a bulk exciton reduced mass of about
$\mu=0.1$~\cite{Miyata2015,Filip2015,Galkowski2016,Yang2017} and dielectric
constant $\epsilon_\mathrm{p}=6.1$, a hydrogenic calculation predicts a bulk
exciton binding energy of 
$E_\mathrm{b} = \mu/\epsilon_\mathrm{p}^2 \times (13.6~\mathrm{eV}) = 37$~meV
and a Bohr radius of 
$a_\mathrm{b}=\epsilon_\mathrm{p}/\mu \times(0.053~\mathrm{nm})=3.2$~nm,
suggesting that the NCs that we study are in the strong to moderate confinement
limits.  (This estimated exciton binding energy is 2--3 times larger than the
most recent experimental measurements~\cite{Miyata2015,Galkoswki2016,Yang2017};
see discussion in Conclusions.)

Calculated absorption spectra are shown in Fig.~\ref{fig:spec}, averaged over
$x$, $y$, and $z$ polarizations.  With increasing size, the spectra become more
structured and the absorption onset energy is reduced due to the quantum
confinement effect. The first excitation energy is indicated by a filled circle
and corresponds to a state that is spectroscopically dark (see further
discussion below), in agreement with previous studies on CsPbBr$_3$
NCs\cite{Tamarat2019, Rossi2020, Rossi2020a, Zhang2020}.  The GW band gap is
indicated by vertical dashed lines. For small NCs, the exciton binding energy is
large, about 1~eV for $L=2.5$~nm ($n=4$). For the largest NCs we study, the
exciton binding energy drops to about 0.3~eV for $L=6.2$~nm ($n=10$).

We now investigate the character of the lowest-lying excitons and the associated
fine structure.  We first review the basic picture of exciton fine structure in
LHPs~\cite{Even2015,Becker2018,Sercel2019JCP,Sercel2019NL}.  The valence band is
primarily composed of Pb 6s and I 5p atomic orbitals (with overall s-like
symmetry) and is doubly degenerate with $m_s = \pm 1/2$.  The conduction band is
strongly split by spin-orbit coupling, and the low-energy band is primarily
composed of Pb 6p$_{j=1/2}$ atomic orbitals and is doubly degenerate with 
$m_j = \pm 1/2$.  This minimal picture suggests three triplet excitons with
$J=1$ and one singlet exciton with $J=0$, all approximately corresponding to an
intra-atomic Pb 6s$\rightarrow$6p transition.  Basic spectroscopic selection
rules suggest that the singlet is dark and the triplet is bright.  Excitons
arising from the higher-energy conduction band with Pb 6$p_{j=3/2}$ character
are at least 1~eV higher in energy, due to the large spin-orbit coupling.  

For a cubic crystal structure, the bright triplet manifold is degenerate and is
higher in energy than the dark singlet exciton $|D\rangle$ due to the
electron-hole exchange interaction.  For a tetragonal crystal structure, which
we use here, the bright triplet degeneracy is broken into a doublet of excitons
$|X\rangle,|Y\rangle$ with polarized absorption along $x$ and $y$ and a singlet
exciton $|Z\rangle$ with polarized absorption along $z$, and their energetic
order depends on the crystal field splitting.  Note that the polarization axes
are aligned parallel to the NC faces, and not along the axes of the tetragonal
unit cell shown in Fig.~\ref{fig:band}(a).

\begin{figure}[t] 
\centering
\includegraphics[scale=0.9]{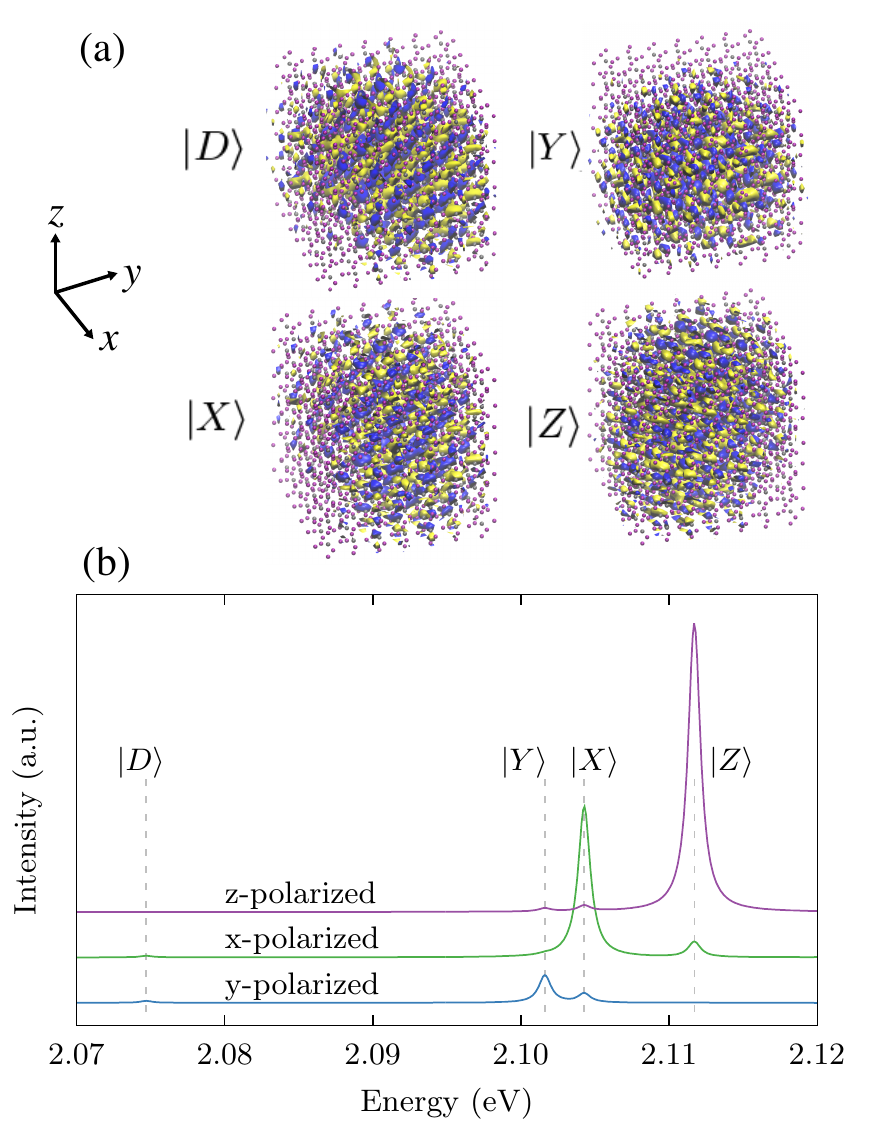} 
\caption{(a) Exciton wavefunction (blue is positive and yellow is negative) of the first four excited states of a $L=4.4$~nm
($n=7$) nanocrystal. The axes are defined along the Pb-I bond, unlike the axes
of the bulk MAPbI$_3$. (b) Absorption spectrum of the same nanocrystal for
$x$-, $y$-, and $z$-polarized light.}
\label{fig:split}
\end{figure}

Our BSE calculations enable direct access to atomistic details of the band-edge
exciton wavefunctions beyond this simplified picture.  As an example, in
Fig.~\ref{fig:split}, we consider a NC with $n=7$ ($L=4.4$~nm).  In
Fig.~\ref{fig:split}(a), we plot the wavefunctions of the lowest four excitons,
\begin{equation}
\Psi^N(\vr_e;\bar{\vr}_h) = \sum_{cv} A_{cv}^N \phi_{c}(\vr_e) \phi_v^*(\bar{\vr}_h) 
\end{equation}
where $\phi_{c/v}(\vr)$ are molecular orbitals, $A_{cv}^{N}$ is the BSE eigenvector,
and we fix the position of the hole $\bar{\vr}_h$ to the center of the NC.
In Fig.~\ref{fig:split}(b), we show the absorption spectrum with the energy of
each transition labeled. Clearly, the lowest energy state is the dark state
$|D\rangle$ and the higher energy states are the three bright states 
$|X\rangle, |Y\rangle, |Z\rangle$, which are labeled according to their
polarized absorption spectra.  We find the $X$/$Y$ energy degeneracy to be
slightly broken along with a weak violation of the polarization selected rules,
which we attribute to the randomly oriented MA cations that lower the crystal
symmetry.  For this relatively small NC, we see a bright-dark splitting of about
30~meV and an intertriplet splitting of about 3--10~meV.

\begin{figure}[t] 
\centering
\includegraphics[scale=0.9]{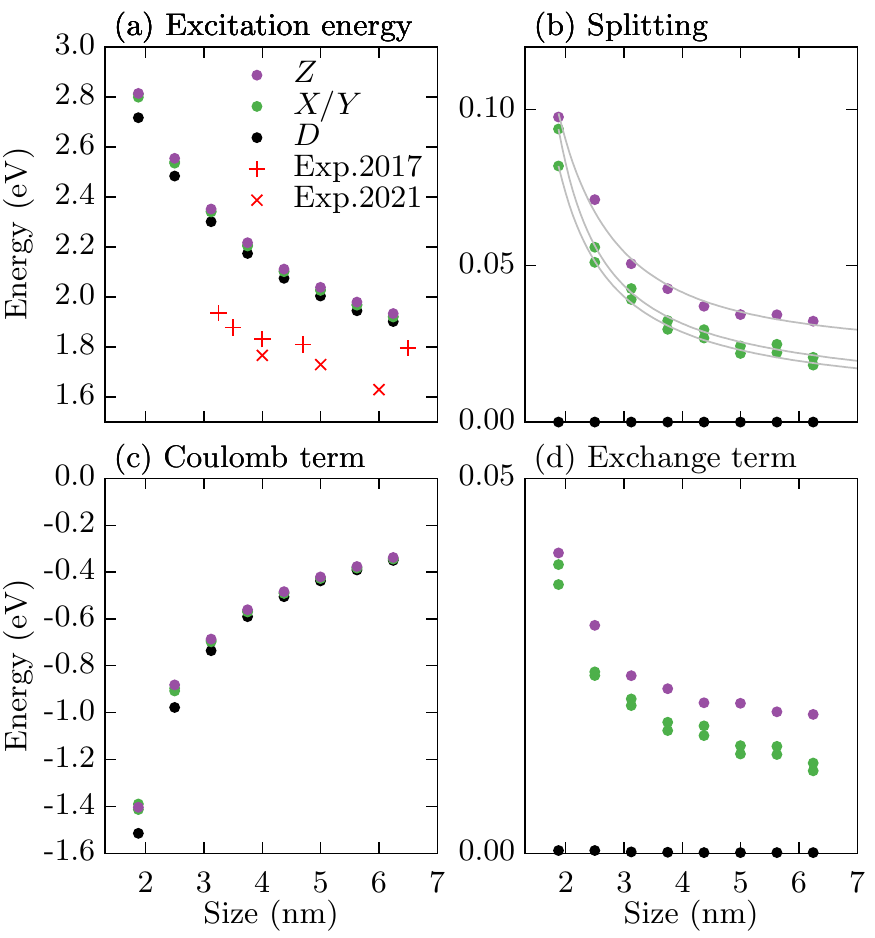} 
\caption{(a) Energy of the first four excited states as a function of the
diameter of the nanocrystal. Experimentally measured photoluminescence values
are from Refs.~\citenum{Anaya2017} (Exp.~2017) and \citenum{Rubino2021}
(Exp.~2021).  (b) Splitting energy of the first four excitation states with
respect to the lowest dark state. Grey lines show the fitting curve for each
excitation state. (c) Coulomb term and (d) exchange term contributing to the
energy of the first four excited states.}
\label{fig:exc}
\end{figure}

Extending this analysis to NCs of other sizes, we find that the energy ordering
is unchanged and that the lowest energy exciton is a dark state, at least up to
$L=6.2$~nm, as shown in Fig.~\ref{fig:exc}(a).  In Fig.~\ref{fig:exc}(b), we
plot the bright-dark splitting of the three bright states as a function of the
size of the NC.  We fit the splitting data to the form
$E_\mathrm{s}(L)=E_\mathrm{s}^\infty+c_1/L+c_2/L^2+c_3/L^3$, consistent with the
binding energy expression in the strong confinement regime~\cite{Sercel2019NL}.
Fitting to our data, we find a nonzero bright-dark splitting of
$E_\mathrm{s}^\infty=4$~meV (for $X/Y$) and 22~meV (for $Z$) in the
$L\rightarrow\infty$ limit.  The large and positive splitting between $Z$ and
$D$ reflects a relatively weak Rashba effect that is insufficient to invert the
ordering~\cite{Sercel2019NL}.  We conclude that, within the approximations made
here, our simulations predict that the lowest exciton is dark for any sized NC. 

Within the GW/BSE approximation, the exciton energies are determined by three
factors: the DFT/GW band  energies (which reflect the local lattice structure,
the carrier confinement, and the dielectric contrast), the electron-hole Coulomb
energy (roughly the exciton binding energy), and the electron-hole exchange
energy.  The latter two quantities are plotted as function of $L$ in
Figs.~\ref{fig:exc}(c),(d).  As expected, we see that the dark exciton has a
vanishing exchange energy and the bright excitons have exchange energies between
40~meV (for small NCs) and 10~meV (for larger NCs).  The exciton binding
energies due to the screened Coulomb interaction are large and vary from 1.5~eV
to 0.3~eV over the same size range.  Interestingly, the dark exciton has a
slightly larger Coulomb interaction than the bright excitons, a feature which
further increases the bright-dark splitting. In fact, for the NC sizes
considered here, we find that the differences in the Coulomb term contribute to
the bright-dark splitting to the same order of magnitude as the exchange energy. 

Finally, we note that the affordable approach taken here allows us to study NCs
that are large enough to be compared directly to experimental results on
MAPbI$_3$ NCs with $L > 3$~nm, which are indicated in Fig.~\ref{fig:exc}(a).
Experimental values come from photoluminescence data of NCs synthesized in the
pores of a patterned SiO$_2$ film~\cite{Anaya2017, Rubino2021}.  We see that our
results overestimate the experimental transition energies by about 0.2~eV. This
level of accuracy is typical even for fully ab initio GW/BSE calculations and is
thus acceptable here, given the additional approximations that were made.  Some
fraction of this discrepancy may also be attributable to a Stokes shift between
absorption and photoluminescence or other exciton-phonon interactions.

\section*{Conclusions}

To summarize, we have performed atomistic GW/BSE calculations of the electronic
and optical properties of realistically sized MAPbI$_3$ nanocrystals, using
parameters determined from ab initio calculations. Within the approximations of
our approach, we find large exciton binding energies of  over 0.3~eV and that
the exciton fine structure is consistent with a spectroscopically dark
lowest-lying exciton. The reasonable agreement between our predicted excitation
energies and experimentally measured photoluminescence energies supports our
methods and findings.

Two approximations in our work are significant and should be revisited for
future improvements. The first is the treatment of the atomic structure:
geometry relaxation may yield a crystal structure that is different from the
bulk tetragonal structure used here. Moreover, the atomic details of the NC
surface, including its passivation and reconstruction, might qualitatively
change our results, especially for small NCs.  The second related approximation
is the neglect of electron-phonon coupling. Bulk LHPs are relatively soft and
anharmonic materials~\cite{Yaffe2017} and strong electron-phonon interactions
have been implicated in their electronic
properties~\cite{Wright2016,Saidi2018,Mayers2018}.  Recent
calculations~\cite{Filip2021,Park2022} have suggested the importance of lattice
screening in determining the exciton properties of LHPs.  For example, the
authors of Ref.~\citenum{Filip2021} find that the typical BSE exciton binding
energy of CsPbX$_3$ is overestimated with respect to experiment by about a
factor of three, and that incorporation of phonon screening partially improves
the agreement.  Given the already high cost of ab initio GW/BSE calculations, we
believe that the approximate approach presented here can serve as a basis for
future studies of atomic structure, surface reconstruction, electron-phonon
coupling, temperature-dependent optical properties, and external dopants.

\section*{Associated Content}
\subsection*{Supporting Information}
The Supporting Information is available free of charge at [publisher inserts link].

\section*{Author Information}
[publisher inserts author information]

\subsection*{Author Contributions}
$^{||}$These authors contributed equally to this manuscript.

\subsection*{Present Addresses}
\textbf{Giulia Biffi} -- Consejo Superior de Investigaciones Científicas, Centro de Física de Materiales, San Sebastián, 20018 Spain
\textbf{Yeongsu Cho} -- Department of Chemical Engineering, Massachussetts Institute of Technology, Cambridge, MA 02139 USA

\subsection*{Notes}
The authors declare no competing financial interest.

\section*{Acknowledgements}
This work has been supported by the H2020-MSCA RISE project COMPASS 691185 (G.B.)
and by the US Air Force Office of Scientific Research
under AFOSR Award No.~FA9550-19-1-0405 (Y.C.).
The authors acknowledge computing resources from the CINECA Calls award under
the ISCRA initiative and from Columbia University’s Shared Research Computing
Facility project, which is supported by NIH Research Facility Improvement Grant
1G20RR030893-01, and associated funds from the New York State Empire State
Development, Division of Science Technology and Innovation (NYSTAR) Contract
C090171, both awarded April 15, 2010.  The Flatiron Institute is a division of
the Simons Foundation.

\providecommand{\noopsort}[1]{}\providecommand{\singleletter}[1]{#1}%
\providecommand{\latin}[1]{#1}
\makeatletter
\providecommand{\doi}
  {\begingroup\let\do\@makeother\dospecials
  \catcode`\{=1 \catcode`\}=2 \doi@aux}
\providecommand{\doi@aux}[1]{\endgroup\texttt{#1}}
\makeatother
\providecommand*\mcitethebibliography{\thebibliography}
\csname @ifundefined\endcsname{endmcitethebibliography}
  {\let\endmcitethebibliography\endthebibliography}{}

\end{document}